# Remarks on the first integral method for solving nonlinear evolution equations

Aparna Saha [1], Benoy Talukdar [1*] Umapada Das [2] and Supriya Chatterjee [3]

[1] Department of Physics, Visva-Bharati University, Santiniketan 731235, India

[2] Department of Physics, Abhedananda College, Sainthia 731234, India

[3] Department of Physics, Bidhannagar College, EB-2, Sector-1, Kolkata 700064, India

[*] Email address : binoy123@bsnl.in

**Abstract.** We point out that use of the first integral method ( J.Phys. A :Math. Gen. **35** (2002) 343 ) for solving nonlinear evolution equations gives only particular solutions of equations that model conservative systems. On the other hand, for dissipative dynamical systems, the method leads to incorrect solutions of the equations.

## 1.Introduction

The problem of finding first integrals of ordinary differential equations (ODEs) was initially considered by Darboux and by Lie [1]. In particular, Darboux showed that one can construct integrating factors and first integrals for a system of polynomial planar differential equations provided there exists a sufficient number of invariant algebraic curves. Sophus Lie, on the other hand, derived a method to find an integrating factor for the first-order differential equation from its admitted symmetries. Then after about a century, Prelle and Singer [2] introduced a semi-algorithmic method to determine the first integral of ODEs. Relatively recently, Feng [3] took recourse to the use of certain results from the theory of commutative algebra to construct an expression for the first integral of Burgers-KdV equation and subsequently employed it to find a solution of the evolution equation. The Burgers-KdV equation is given by

$$u_t + \alpha u u_x + \beta u_{xx} + \gamma u_{xxx} = 0, \quad u = u(x,t) \qquad (1)$$

with the parameters $\alpha, \beta, \gamma \neq 0$. Feng [3] converted the above partial differential equation into an ordinary differential equation

$$\phi''(\xi) - r\phi'(\xi) - a\phi^2(\xi) - b\phi(\xi) - d = 0 \qquad (2)$$

using the travelling coordinate $\xi = x - vt$ and a change of variable $u(x,t) = \phi(\xi)$. Here $v$ stands for the non-zero translational wave velocity and $d$, a constant of integration. Also $r = -\beta/\gamma, a = -\alpha/2\gamma, b = v/\gamma$. Equation (2) is equivalent to a system of first-order equations

$$X'(\xi) = Y(\xi) \text{ and } Y'(\xi) = rY(\xi) + aX^2(\xi) + bX(\xi) + d \qquad (3)$$

where $X(\xi) = \phi(\xi)$ and $Y(\xi) = \phi'(\xi)$. The primes in (3) denote differentiation with respect to $\xi$. The planar equations in (3) can be combined to write

$$\frac{dY^2}{dX} - 2rY = 2aX^2 + 2bX^2 + 2d . \qquad (4)$$

The nonlinear equation in (4) cannot be solved analytically to construct a first integral of (2). The source of trouble is the second term on the left. Note that this term has its origin in the dissipative-like term in (2). In view of this, Feng introduced an ansatz to write the first integral for (2) and, as we already noted, used it to provide a solution of the evolution equation. The method introduced by Feng is often called the first integral method. In this context we note that one can avoid use of the so-called first integral method for equations which lead to solvable planar equations. Despite that, the first integral method has been extensively used to treat such integrable systems. The term integrability is used here not in its technical sense, but we call all equations integrable if the corresponding planar systems can be solved analytically. In the present work we shall demonstrate that even for solvable equations the method of Feng, does not give their general solutions and for unsolvable problems the method leads to first integrals which are not constants on the solution curves. Thus all results for dissipative dynamical systems obtained by the method of Feng are incorrect.

We consider in section 2 two integrable systems modeled by modified Benjamin-Bona-Mahony and by Fornberg - Whitham equations. We find general solutions of the equations by means of direct integration. In each case the constructed solution is characterized by the Hamiltonian ($H$) of the system such that for a particular value of $H$ our general result yields the special solution obtained by the first integral method. The same value of $H$ when substituted in the expression for the first integral found by us leads to the corresponding result appearing in the so-called first integral method. We then find Lagrangian representations of the systems. In section 3 we deal with the Burgers-KdV equation and demonstrate that the ansatz of the first integral as found by Feng [3] is not constant on the solution curve so as to provide us with a correct solution of the problem. We further show that (2) does not represent a Lagrangian system. The results presented by us confirm that only Lagrangian system of equations can be treated by the first integral method to construct their particular solutions and for non-Lagrangian systems use of the method for which it was originally designed leads to incorrect results. Finally, in section 4 we summarize our outlook on the present work and make some concluding remarks.

## 2. Integrable equations

As a solvable model we first consider the modified Benjamin-Bona-Mahony (MBBM) equation

$$u_t + u_x + a_1 u^2 u_x + b_1 u_{xxt} = 0. \qquad (5)$$

For (5), equations similar to those in (2) and (4) are given by

$$\phi'' = \alpha_1 \phi + \beta_1 \phi^3 \qquad (6)$$

and

$$\frac{dY^2}{dX} - 2\alpha_1 X - 2\beta_1 X^3 = 0 \qquad (7)$$

respectively. Here $\alpha_1 = (1-v)/b_1 v$ and $\beta_1 = a_1/3 b_1 v$. On integration (7) yields

$$Y = -\sqrt{\frac{\beta_1}{2}} \sqrt{X^4 + 2\alpha_1 X^2 / \beta_1 + 2H / \beta_1} \qquad (8)$$

where the constant of integration $H$ stands for the Hamiltonian of the system modeled by the MBBM equation since (6) does not involve the time parameter explicitly. Remembering that $Y = X'(\xi) = \phi'(\xi)$ we can solve (8) to get

$$\phi(\xi) = -i\sqrt{\frac{2H}{\alpha_-}} sn\left(\sqrt{\frac{\alpha_-}{2}} i(\xi + \xi_o), \frac{\alpha_+}{\alpha_-}\right) \qquad (9)$$

where $\xi_o$ is a constant of integration and $sn(\omega, \rho)$ stands for the Jacobi sine function [4]. Here

$$\alpha_\pm = \alpha_1 \pm \sqrt{\alpha_1^2 - 2\beta_1 H} . \qquad (10)$$

For $H = \alpha_1^2 / 2\beta_1$, (9) gives

$$\phi(\xi) = -\sqrt{\frac{\alpha_1}{\beta_1}} \tan\left(\sqrt{\frac{\alpha_1}{2}}(\xi + \xi_o)\right). \tag{11}$$

This result for the particular value of $H$ as indicated above was obtained by Taskan and Bekir [5] by the use of first integral method. The same value of $H$ when substituted in (8) gives the first integral

$$Y = -\sqrt{\frac{\beta_1}{2}}\left(X^2 + \frac{\alpha_1}{\beta_1}\right) \tag{12}$$

on the basis of which the solution (11) was derived in ref. 5. One expects to obtain (6) by differentiating (12) with respect to $\xi$. But this is not true. However, one can easily verify that the square of (12) when differentiated with respect to $\xi$ gives (6).

In close analogy with the form of Hamilton's equations in classical mechanics, a system of planar equations

$$X' = f(X,Y), \ Y' = g(X,Y), \ X = X(\xi)$$

$$Y = Y(\xi), \ X' = \frac{dX}{d\xi}, \ Y' = \frac{dY}{d\xi} \tag{13}$$

is called a Hamiltonian system provided there exists a function $H(X,Y)$ such that

$$f = \frac{\partial H}{\partial Y}, \ g = -\frac{\partial H}{\partial X}. \tag{14}$$

Then $H$ is called the Hamiltonian of the system. A necessary and sufficient condition for (14) to be Hamiltonian is that

$$\frac{\partial f}{\partial X} + \frac{\partial g}{\partial Y} = 0. \tag{15}$$

It is straightforward to verify that the Hamiltonian $H$ calculated from (8) as a function of $X$ and $Y$ when substituted in Eq.(14) will lead to the planar equations corresponding to (6). This establishes that Eq. (6), or equivalently MBBM equation, models a Hamiltonian or conservative system. The Lagrangian for (6) is given by

$$L(\phi, \phi', \xi) = \frac{1}{2}\phi'^2 + \frac{1}{4}\beta_1\phi^4 + \frac{1}{2}\alpha_1\phi^2 \tag{16}$$

which when substituted in the Euler-Lagrange equation gives back the equation.

The other integrable system of our interest is modeled by the Fornberg-Witham equation

$$u_t - u_{xxt} + u_x + uu_x = uu_{xxx} + 3u_x u_{xx}. \tag{17}$$

In the traveling coordinate, (17) can be written as

$$\phi''(\phi - v) + (v-1)\phi - \frac{1}{2}\phi^2 + \phi'^2 = k \tag{18}$$

where $k$ is a constant of integration. In the planar equations corresponding to (!8) there appears a singular line $x = v$. We introduce a transformation $d\eta = (x - v)d\xi$ in order to avoid this line of singularity and thus write the modified system of equations

$$\frac{dX}{d\eta} = (X - v)Y \tag{19a}$$

and

$$\frac{dY}{d\eta} = (1 - v)X + \frac{1}{2}X^2 - Y^2 + k. \tag{19b}$$

Equations in (19) can be combined to write

$$\frac{dY^2}{dX} + \frac{2}{X-v}Y^2 = \frac{2}{X-v}\frac{2k + 2(1-v)X + X^2}{X-v}. \tag{20}$$

Solving the above first-order inhomogeneous differential equation in $Y^2$ we get

$$Y = \sqrt{\frac{3X^4 + (8+12v)X^3 + 12(v^2 + k - v)X^2 - 24vkX - 12H}{12(X-v)^2}}. \tag{21}$$

We have found that the numerator of the expression under the radical in (21) becomes divisible by the denominator provided

$$H = \frac{1}{12}(3v^4 - 4v^3 - 12v^2k). \tag{22}$$

From (21) and (22) we thus write

$$Y = \frac{\sqrt{X(8 + 3X) + v(4 - 6X) + 12k - 3v^2}}{2\sqrt{3}}. \tag{23}$$

Remembering that $Y = \frac{dX}{d\xi}$ and $X = \phi(\xi)$ we solve (23) such that

$$\phi(\xi) = \frac{1}{6}e^{-(\xi+\xi_0)/2}\left(16 - 36v + 18v^2 - 36k - (8 - 6v)e^{(\xi+\xi_0)/2} + e^{(\xi+\xi_0)}\right) \tag{24}$$

where $\xi_0$ is again a constant of integration. Further a special value of $k$ chosen as

$$k = \frac{1}{18}(8 - 18v + 9v^2) \tag{25}$$

reduces (24) to the particular solution

$$\phi(\xi) = v - \frac{1}{3}(4 - e^{-(\xi-\xi_0)/2}) \tag{26}$$

of the Fornberg-Whitham equation obtained by Aslan [6] by the use of first integral method. The corresponding relation for the first integral as found from (23) and (25) is given by

$$Y = \frac{1}{2}X + \frac{1}{6}(4 - 3v).  \tag{27}$$

The first integral in (27) has a worse pathological property than the result in (12). For example, we can never obtain the Fornberg-Witham equation from (27).

As with the MBBM equation, the Fornberg-Witham equation also forms a Lagrangian/Hamiltonian system with the Lagrangian given by

$$L = \frac{1}{2}(\phi - v)^2 \phi'' + \frac{\phi^4}{8} + \frac{1}{6}(2 - 3v)\phi^3 + \frac{1}{2}(v^2 - v + k)\phi^2 - vk\phi.  \tag{28}$$

### 3. Non-integrable planar equations and first integral method

Our results of the previous section clearly show that for integrable equations use of the first integral method can provide us with particular solutions only. Moreover, the relation of the first integral as found by the method tends to pose some conceptual problem. The Burgers-KdV equation is not integrable. It, therefore, remains an interesting curiosity to examine if the relation of the first integral [3]

$$Y(\xi) = \frac{2r}{5}X(\xi) + \frac{2br}{5a} \pm \sqrt{\frac{2}{3a^2}(aX(\xi) + b)^3}  \tag{29}$$

on differentiation with respect to $\xi$ yields (2). We have checked that it is not possible. Thus the solution of the Burgers-KdV equation calculated from (29) is incorrect. In the following we use Helmholtz theorem [1] of the inverse variational problem to provide adequate justification for why (29) does not represent a first integral.

Let $P[w] = P[x, w^n] \in A^r$ be an $r$-tuple differentiable function. The Frechet derivative of $P$ is the differential operator $D_P : A^q \to A^r$ defined by

$$D_P(Q) = \frac{d}{d\varepsilon}\bigg|_{\varepsilon=0} P[w + \varepsilon Q[w]]  \tag{30}$$

for any $Q \in A^r$. If $D = \sum_J P_J[w]D_J$, $P_j \in A$ is a differential operator, its adjoint $D^*$ is defined by

$$D^* = \sum_J (-D_J)P_j.  \tag{31}$$

Helmholtz theorem asserts that any nonlinear evolution equation $w_t = P[w]$ will have a Lagrangian representation only if $D_P$ is self-adjoint i.e. $D_p = D_P^*$. To see whether the Burgers-KdV equation will admit such a representation we proceed as follows

.A single evolution equation is never an Euler-Lagrange expression. One common trick to put an evolution equation into variational form is to replace $u(x,t)$ by a potential function $w(x,t)$ such that

$$w(x,t) = \int_x^\infty dy\, u(y,t). \tag{32}$$

From (1) and (32) we can write

$$P[w] = \frac{\alpha}{2} w_x^2 - \beta w_{xx} - \gamma w_{xxx}. \tag{33}$$

We now make use of (30) to compute the Frechet derivative of $P[w]$ and obtain

$$D_p(Q) = \alpha w_x D_x Q - \beta D_x^2 Q - \gamma D_x^3 Q. \tag{34}$$

Using (31) it is easy to verify that $D_p \neq D_P^*$. Thus Eq.(1) and /or equivalently Eq. (2) does not have a Lagrangian representation to follow from the action principle. Thus the Burgers-KdV equation models a dissipative system such that its first integral will have explicit $\xi$ dependence. But the relation in (29) does not exhibit this behavior. Consequently, (29) is not at all a first integral of the Burgers-KdV equation. The conclusion drawn here is true for any non-integrable/dissipative system.

**5. Concluding remarks**

The first integral method [3] has been extensively used in the literature with a view to provide new exact travelling wave solutions of physically interesting nonlinear evolution equations. The problems treated so far can broadly be divided into two classes – the first one deals with equations for which it is rather trivial to find first integrals while the second one is concerned with equations which do not appear to admit first integrals.

In this work we have first chosen to work with two integrable equations belonging to class I and converted them to Hamiltonian form. We have solved these equations in terms of Jacobi elliptic and exponential functions. For some particular values of the Hamiltonian function our solutions go over to those given by the use of the so-called first integral method [5,6]. The chosen values of the Hamiltonian also reproduce the ansatz for the first integrals as introduced in refs. 5 and 6. Unfortunately, these limiting results are not constants on the solution curves. This awkward analytical constraint appears to be more prominent in the original work of Feng [3]. We, therefore, conclude by noting that for integrable nonlinear evolution equations use of the first integral method gives only particular solutions and for non-integrable equations the method leads to incorrect solutions. We attribute the reasons for this to different variational properties of the integrable and non-integrable sustems.

**Acknowledgement**. One of the authors (BT) likes to thank Dr. Debabrata Pal for his kind interest in this work.